\documentstyle[preprint,tighten,prb,aps]{revtex} 
\begin{document}
\draft
%\preprint{IF--UFF 03/95}
%\twocolumn[\hsize\textwidth\columnwidth\hsize\csname @twocolumnfalse\endcsname
\title{Bond-charge Interaction in the Extended Hubbard Chain}

\author{Daisy M.~Luz\cite{emaild} and Raimundo R. dos Santos \cite{email}}

\address{Instituto de F\'\i sica, 
         Universidade Federal Fluminense,
         24020-150 Niter\'oi, Rio de Janeiro, Brazil}
\date{\today}
\maketitle

\begin{abstract}
We study the effects of bond-charge interaction (or correlated
hopping) on the properties of the extended ({\it i.e.,} with both
on-site ($U$) and nearest-neighbor ($V$) repulsions) Hubbard model 
in one dimension at half-filling. Energy gaps and correlation
functions are calculated by Lanczos diagonalization on finite
systems.
We find that, irrespective of the sign of the bond-charge 
interaction, $X$, the charge--density-wave (CDW) state is more 
robust than the spin--density-wave (SDW) state.
A small bond-charge interaction term is enough to make the
differences between the CDW and SDW correlation functions much
less dramatic than when $X=0$.
For $X=t$ and fixed $V<2t$ ($t$ is the uncorrelated hopping 
integral), there is an intermediate phase between a charge ordered phase
and a phase corresponding to singly-occupied sites, 
the nature of which we clarify: it is characterized by a 
succession of critical 
points, each of which corresponding to a different density of
doubly-occupied sites. 
We also find an unusual slowly decaying staggered
spin-density correlation function, which is suggestive of 
some degree of ordering.
No enhancement of pairing correlations was found for any $X$ in the
range examined. 
\end{abstract}
%\bigskip
%\vskip2pc]

\section{INTRODUCTION}
\label{sec:intro}

Over recent years  bond-charge interactions between fermions 
have been invoked to explain a variety of phenomena, such as 
lattice stiffening in polyacetylene,\cite{Kivelson87,Gammel88}
and high-temperature superconductivity.\cite{Hirsch89,Essler92,Appel93}
This kind of interaction had already appeared in the study of magnetism 
of narrow $d$-band electrons as discussed by Hubbard:\cite{Hubbard63}
the Coulomb interaction matrix elements are expressed in terms of Wannier states $|i\rangle$ localized at sites $i$,
giving rise not only to the well-known on-site repulsion, 
$U=\langle ii|{1\over r}|ii\rangle$, and nearest-neighbor repulsion
$V=\langle ij|1/r|ij\rangle$, but also to $X=\langle ii|{1\over r}|ij\rangle$.
The Hamiltonian then becomes
\begin{eqnarray}
%\begin{equation}
\label{Ham}
{\cal H}&=&-\sum_{\langle ij\rangle,\ \sigma}
\left[t-X(n_{i-\sigma}+n_{j-\sigma})\right]\ 
\left(c_{i\sigma}^{^{\dagger}} c_{j\sigma}+\text{H.c.}\right)
\nonumber\\ 
&&+ U\sum_i n_{i\uparrow}n_{i\downarrow}
+V \sum_{\langle ij\rangle,\ \sigma\sigma'}n_{i\sigma}n_{j\sigma'}\ ,
%\end{equation}
\end{eqnarray}
where $\langle ij\rangle$ stands for nearest-neighbor sites,
$c_{i\sigma}^{^\dagger}$ ($c_{i\sigma}$) creates (annihilates) a fermion
at site $i$ in the spin state $\sigma=\uparrow\ {\rm or}\ \downarrow$, and
$n_i=n_{i\uparrow}+n_{i\downarrow}$, with 
$n_{i\sigma}=c_{i\sigma}^{^\dagger}c_{i\sigma}$. 
The second-quantized form of the interaction justifies calling this extra term a bond-charge interaction, also referred to as correlated-- or 
density-dependent--hopping: it favors (hinders) hopping involving sites 
with non-zero charge if $X<0$ ($X>0$).
While in the original problem the parameters were such that $U\gg V\gg X$,
in more recent applications\cite{Kivelson87} one may have 
$U\gg V\sim X$, or even $U\sim V\sim X$, such as in 
an effective model for the CuO$_2$ planes in high-temperature 
superconductors.\cite{Appel93} 

For the case of on-site repulsion only (i.e., $X=V=0;\ U\neq 0$) 
an exact solution in one dimension based on the Bethe ansatz 
has been known for some time.\cite{Lieb&Wu68} 
For half-filled band and for strong repulsion, the ground state 
is reminiscent of a N\'eel state, but differing in a fundamental way: 
the staggered spin correlations are {\em critical}, corresponding to an
algebraic spatial decay; 
this is commonly referred to as spin-density-wave (SDW) state.
When nearest-neighbor repulsion ($V$) is included, no exact solution
has been found, even in one dimension.
For nearest-neighbor repulsion in the absence of both hopping and
on-site repulsion, the ground state displays charge ordering (CO), 
since it is more favorable for the electrons to doubly-occupy alternate
sites, leaving their neighbors empty. 
As the hopping is switched on, quantum fluctuations disturb this ``static''
ordering, turning into a charge-density-wave (CDW) state. 
Different approximations have been used in order to obtain the phase
diagram for intermediate values of both $U$ and $V$;
the picture that emerges is that there is a CDW--SDW phase transition 
at zero temperature, along a critical curve slightly 
displaced from the strong-coupling prediction $V_c=U/2.$\cite{Cabib75,F&S84a,F&S84b,Hirsch84,Hirsch85b,Cannon90}
%Further, the transition is of first order for $U>U_t$ (see, {\it e.g.,} 
%Ref.\ \onlinecite{Hirsch85b} for a qualitative `droplet' picture in the 
%strong coupling regime) and of second order for $U<U_t$; $U_t$ is then a 
%tricritical point. 

The information on the effects brought about by the bond-charge 
interaction in the presence of {\it repulsive} on-site and inter-site
interactions (i.e., $U,V>0$) is somewhat limited. (For either $U$ or
$V$ attractive, see e.g, Refs.\ \onlinecite{Essler92} and
\onlinecite{Micnas90}).
On the one hand, mean-field--type calculations\cite{Lara90} predict 
that bond-charge
interaction gives rise to superconductive pairing for $4zX^2>UV$ 
($z$ is the coordination number); though very interesting, the 
validity of this result in one and two dimensions should be 
questioned. 
On the other hand, when $X=t$ the total number of doubly occupied sites,
$N_2\equiv\sum_in_{i\uparrow}n_{i\downarrow}$,
becomes a conserved quantity, allowing exact statements to be 
made\cite{Strack93,Ovchinnikov934,Aligia95}
for the model defined by (\ref{Ham}) at half-filling ($\langle n\rangle=1$),
and for any lattice dimensionality $d$: 
(i) For $U<2zV-z\ {\rm max}(2t,V)$, the ground state corresponds to
a non-magnetic charge ordered
metal.\cite{Ovchinnikov934,Aligia95}
(ii) For $U>z\ {\rm max}(2t,V)$, the ground state corresponds to a
paramagnetic insulator, with exactly one fermion at every site\cite{Strack93} [singly occupied (SO) sites], leading to a very
high degeneracy with respect to the total spin value.\cite{Ovchinnikov934}
Further, Ovchinnikov\cite{Ovchinnikov934} also argues that as
$U$ decreases below the SO boundary, $U=4t$ when $V<2t$, the appearance of
doubly-occupied sites (henceforth referred to as the DO region) is favored.
Also for $X=t$, but in one dimension with $V\equiv 0$, one finds 
three regimes, differing by the allowed site occupancy: 
(i) no doubly occupied sites [no empty sites] for 
$\langle n\rangle<1$ [$\langle n\rangle>1$]; 
(ii) no singly occupied sites; (iii) all 
possibilities.\cite{Ovchinnikov934,Arrachea94} 

In view of the wide range of applications of bond-charge interaction, 
a clear understanding of the zero-temperature phase diagram is 
clearly in order.
In particular, the behavior with $X$ of the boundary between the 
CDW and the SDW phases is of
interest, together with any indication of superconducting correlations
becoming dominant.
When $X=t$ most of the definite statements made in relation
to the intermediate (DO) phase concern the location of the boundaries with 
the CO and SO phases, while a precise understanding of its nature and of its
properties is still lacking.  
With this in mind, here we consider the half-filled model in one 
dimension and discuss the phase diagram; from now on, energies will 
be expressed in units of the usual hopping, $t$. 

The layout of the paper is as follows. In Sec.\ \ref{sec:calcs} we
briefly outline the procedure used to determine the phase-transition
boundaries and several correlation functions, which will be useful
in interpreting the nature of the phases involved.
In Sec.~\ref{sec:Xn1} we present our results for the case without
double occupancy conservation, i.e., for $-1\leq X<1$;
the case $X=1$ is discussed in Sec.~\ref{sec:X1}.
And finally, Sec.~\ref{sec:concl} summarizes our findings.

\section{ENERGY GAP AND CORRELATION FUNCTIONS}
\label{sec:calcs}

The Hamiltonian (\ref{Ham}) for chains with $N_s$ sites and $N_e=N_s$
fermions (half-filled band) is diagonalized
through the Lanczos algorithm;\cite{Roomany80,Gagliano86,Paige76} 
periodic (anti-periodic) boundary conditions are used for $N_s=4,8,12$
($N_s=6,10$).\cite{F&S84b}
We obtain the lowest eigenvalues and corresponding eigenvectors,
from which the energy gap, defined as
\begin{equation}
\Delta=E_1-E_0
\label{gap}
\end{equation}
is calculated, where $E_0$ and $E_1$ are the two lowest energy eigenvalues.
The gap defined above is at  constant  number of particles, 
unlike the charge gap $\Delta_c$ used in Ref.\ \onlinecite{F&S84b},
which is related to a finite-difference approximant to the inverse  
compressibility.
At zero temperature, and in the thermodynamic limit (TDL), 
a phase transition is marked 
by a change in the ground state; one therefore expects the energy difference between
the two lowest states to vanish at the transition point. 
While in many quantum systems the energy gap, defined as in Eq.\ (\ref{gap}),
only vanishes at the transition in the TDL,\cite{dS81} 
in the present case $\Delta$ vanishes for finite-sized systems,
and it will be used to estimate the location of CDW-SDW transition
points. 
For a given $X$, we fix $V$ and determine $U_c$, the value of $U$ where $\Delta$ vanishes.  
In principle, the location of the transition point, $U_c(V,X)$, depends
on the system size, $N_s$, and we have to perform extrapolations towards 
$N_s\to\infty$.

As a test of this procedure, we show in Fig.\ \ref{pd1} our results 
for the CDW-SDW transition line for $X=0$.
We have found excellent agreement with other
estimates,\cite{F&S84b,Hirsch84,Hirsch85b,Cannon90} including the deviation
of the phase boundary from the strong coupling prediction $V_c=U/2$ for
$U\gtrsim 1$, confirming that the gap defined by Eq.\ (\ref{gap}) is indeed
appropriate. 

In addition to the energy gap, the nature of the different phases
is probed by the staggered spin-spin correlation function,
\begin{equation}
\label{sdw}
{\cal F}_S(\ell)=(-1)^\ell\langle m_i m_{i+\ell}\rangle,
\end{equation} 
by the charge-density correlation function,
\begin{equation}
\label{cdw}
{\cal F}_C(\ell)={1\over 4}\langle n_i n_{i+\ell}\rangle,
\end{equation} 
by the singlet superconductor correlation function,
\begin{equation}
\label{ss}
{\cal F}_{SS}(\ell)=\langle c_{i+\ell\downarrow} c_{i+\ell\uparrow}
c^\dagger_{i\uparrow} c^\dagger_{i\downarrow}\rangle,
\end{equation} 
and by the triplet superconductor correlation function,
\begin{equation}
\label{ts}
{\cal F}_{TS}(\ell)=\langle c_{i+\ell\downarrow} c_{i+\ell+1\uparrow}
c^\dagger_{i+1\uparrow} c^\dagger_{i\downarrow}\rangle,
\end{equation}
where 
\begin{equation}
\label{m}
m_i=n_{i\uparrow}-n_{i\downarrow},
\end{equation}
and
\begin{equation}
\label{n}
n_i=n_{i\uparrow}+n_{i\downarrow}.
\end{equation}
One should note that with the above definitions the $\ell\to\infty$ limit 
of the charge correlations is 1/4, whereas all others go to zero.

As further tests of our calculations, the dotted lines in Fig. \ref{corr2}
represent the behavior of spin and charge correlations in the absence of 
bond-charge interaction.
Figure \ref{corr2}(a) shows that in the CDW phase, charge correlations
are slowly decaying with the distance, while spin correlations are rapidly
decaying, indicating that the former are dominating. 
In the SDW phase the roles are inverted, as shown in Fig.\ \ref{corr2}(b).
As is well known, the system is far from a superconducting instability
in this case, which is reflected by the singlet and triplet correlations
falling off with the distance much faster than their spin or charge couterparts.
We have also tested for any crucial size dependence of the above correlation
functions, and found that the results for $N_s=12$ are hardly different 
from those for $N_s=10$ or 8. 
These procedures will now be used in the analysis of the role played
by bond-charge interaction.

\section{RESULTS FOR $-1\leq X<1$}
\label{sec:Xn1}

We now discuss the effects of correlated hopping on the phase diagram
and correlation functions. Using the procedure outlined in 
Sec.\ \ref{sec:calcs}, the vanishing of the gap locates a phase
transition point.
Figure \ref{fphased} shows constant-$X$ sections of the phase
diagram.  
For non-zero $X$, the CDW-SDW phase boundary is displaced  to
the right of that corresponding to $X=0$; that is, the CDW region grows 
with $|X|$, 
at the expense of the SDW region; see Fig.\ \ref{fphased}.
The case $X=\pm 0.5$ illustrates that
while for large $U$ (and $V$) the location of the phase boundary independs 
on the sign of $X$, this is not so for $U\lesssim2.5$: the CDW region is 
larger for $X>0$ than for $X<0$. 
A rough picture can be used to explain this sign dependence, as follows.
The dominant contribution to the ground state in the CDW phase is from a
charge-ordered state 
($|\ldots\uparrow\downarrow\ 0\ \uparrow\downarrow\ 
0\ \uparrow\downarrow\ldots\rangle$).
For small $V$, a spin resonating between two sites would typically lower the
energy by $\sim U$ for breaking the ``pair'', and add $\sim X$ due to 
the correlated hopping (in order to stress the role of the bond-charge energy,
we do not consider the $t$-hopping). 
Since one can think of hopping-induced ``pair''-breaking as the mechanism 
by which the CDW state becomes an SDW state, a smaller $U_c$ is needed when $X<0$.

In Fig.\ \ref{corr2}(a) we show the effect of bond-charge interaction
on the correlation functions in the CDW region, by comparing the results 
for $X=0$ (dotted lines) with the corresponding ones for $X=-0.1$ (solid lines).
Charge correlations, which are slowly decaying in the absence of bond-charge
interaction, become strongly damped when $X\neq0$.
Spin correlations, on the contrary, are enhanced when bond-charge 
interaction is present. Turning into the SDW phase, we see from the data
shown in Fig.\ \ref{corr2}(b) that the behavior is exactly the opposite of
that in the CDW phase: charge correlations are enhanced, while spin 
correlations are depressed. 
Though with somewhat smaller amplitudes of charge oscillations, the 
behavior is qualitatively the same as $V$ is decreased.
Therefore, the differences between CDW and SDW phases, which are very 
pronounced when $X=0$, become quite smooth already for a small value of $X$.
Since when $X=0$ the CDW-SDW transition is expected to be of first order
for $U>1.5$ (see e.g., Ref.\ \onlinecite{Cannon90}), this may be indicative 
that  bond-charge interaction weakens the first order character 
of this transition. 

This fact has other consequences. In a standard strong-coupling
perturbation theory analysis,\cite{Hirsch84} the critical curve would
be given by
\begin{equation}
V_c={U\over 2}+1.545\ {(t-X)^2\over U},\\
\label{vc}
\end{equation}
which would indicate a growth of the SDW phase into the CDW, 
contrary to the behavior shown in Fig.\ \ref{fphased}. 
The results displayed in Fig.\ \ref{corr2} reflect the fact
that the contribution of SDW-like states to the ground state
in the CDW region is much more important in the presence of 
bond-charge interaction than when $X=0$; 
a similar statement holds for the contribution of
CDW-like states in the SDW region.   
Since the standard strong coupling analysis neglects  
these contributions altogether, quantum fluctuations other than those
present in Eq.\ (\ref{vc}) could account for the above discrepancy.

As far as superconducting correlations
are concerned, they are not enhanced by the presence of bond-charge
interaction in either phase, meaning that no tendency towards
pairing has been detected at half-filling.

\section{RESULTS FOR $X=1$}
\label{sec:X1}
The case $X=1$ shows remarkable features due to the 
conservation of the number of  doubly-occupied sites.\cite{Strack93}
This regime can therefore be considered as quasi-classical, in the sense
that fluctuations brought about by the uncorrelated hopping term
are limited by the conservation of the number of doubly-occupied sites.
%That is, the fermions are nearly pinned.
Computationally, this considerably restricts the Hilbert space, 
accelerating the Lanczos algorithm; we were able to consider
lattices up to $N_s=24$ in this case.

The energy spectrum in the intermediate region deserves a closer look. 
One finds three different regimes as $U$ is varied.
For $U<U_{c_1}(N_s,V)$, the ground state corresponds to full charge
ordering (CO), with energy per fermion $E_{CO}/N_s=U/2$.
At the other extreme, $U>U_{c_2}(N_s,V)$, the ground state
corresponds to singly-occupied (SO) sites, with energy $E_{SO}/N_s=V$.
In-between those limits, the number of doubly occupied (DO) sites plays 
a dominant role in selecting the ground states.
The calculated energy (per fermion) of the lowest state with one 
doubly occupied (DO) site, for a given system size, corresponds 
to a straight line with slope $1/N_s$. 
Since the regime $X=1$ can be considered as quasi-classical, this is the
strong coupling result, where the energy cost of having one doubly 
occupied site ($N_2=1$) is $U$.
Similar analyses for the cases $N_2=2, 3,\ldots$, indicate
that the slopes of the lowest energy levels are given by $N_2/N_s$, if 
$N_2\leq ({N_s\over 2}-1)$; 
recall that $N_2={N_s\over 2}$ corresponds to having all sites doubly occupied,
which is the CO state. 
It is interesting to note that the dominant contributions to the ground state 
comes from states where the DO sites are farthest apart, 
so they are evenly distributed throughout the chain, and singly occupied
sites surround both sides of the DO ones. 
For instance, when $N_2=4$ on a 12-site ring, the dominant contributions to the ground state are
$$%\begin{equation}
|0\ \uparrow\ \ \uparrow\downarrow\ \ \downarrow\ 0\ \uparrow\downarrow\ 0\ 
\ \uparrow\ \ \uparrow\downarrow\ \ \downarrow\ 0\ \uparrow\downarrow\rangle +
{\rm C.p.,}\nonumber\\
$$%\end{equation}
where C.p. stands for circular permutations.

When the lowest energies corresponding to different number of DO sites are
compared, several regimes can be clearly distinguished. 
For a given system size, as $U$ decreases below $U_{c_2}$, the lowest states
correspond in succession to one, two, three,\ldots doubly occupied sites. 
On the other hand, since the energy densities extrapolate to horizontal 
lines in the thermodynamic limit (see the discussion above), 
one may be misled to think that all energies should merge in that limit,
amounting to a macroscopic degeneracy of states.
The proper way to analyse the DO region is therefore in terms of 
{\em densities} of DO sites, in a situation analogous to that of occupancy. 
That is, for a given ratio $N_2/N_s$ we calculate the lowest energies for
chains with $N\cdot N_2$ DO sites out of $N\cdot N_s$ sites, with 
$N=1$, 2, 3, and so forth. 
The energy levels thus obtained extrapolate 
as $N_s\to \infty$ to the ones shown in Fig.\ \ref{energies3}.
Though for finite systems the densities do not vary continuously, 
the following trend can be inferred from Fig.\ \ref{energies3}:
Below $U_{c_2}$ the ground state suffers a succession of transitions to states
with continuously increasing density of DO sites. In a Renormalization Group
language, for fixed $V$, the DO region consists of a line of fixed points
between the CO and the SO regions; each fixed point corresponds to a density
of DO sites. As $V$ is varied, this line of fixed points is displaced 
accordingly, and we may view the DO region in the phase diagram of 
Fig.\ \ref{PDX1} as a critical region made up of fixed points.   

Another remarkable feature of the $X=1$ region is that states with 
$S^z\equiv\sum_i\langle m_i\rangle=N_s-2N_2$, corresponding to the maximum
$S^z$ compatible with the number of DO sites, are degenerate with those
with $S^z=0$. 
In Ref.\ \onlinecite{Ovchinnikov934} it was shown that if one assumes the
stability of the ferromagnetic state (similarly to the Nagaoka\cite{Nagaoka}
problem), then there is a full degeneracy in $S^z$ at the transition point
$V=0,\ U=4$. Our results show that this degeneracy is partially lifted
for $V\neq0$.
    
Further insight is obtained by examining the 
correlation functions defined by Eqs.\ (\ref{sdw})-(\ref{ts}).
In the CO region, the charge correlation function alternates between
1/2 and 0, without any decaying with the distance; similarly, the 
spin density correlations vanishes identically, while superconducting
correlations are zero for $\ell\neq0$. This clearly confirms the static
picture.
In the SO region, the charge-density correlation function is uniform,
sticking to the
value 1/4, the usual asymptotic value corresponding to one fermion per site. 
Since all singly-occupied states are degenerate, irrespective of $S^z$ and 
of the spin arrangement for a given $S^z$, the specific form of spin-density
correlation function depends on the ground state one is considering, though
the magnitude is always 1.
Again, no relevant superconducting correlations were observed.

We now turn to the analysis of correlations in the more interesting DO region.
First we consider the case of one DO site.
While charge correlations attain their limiting value 1/4 for any $\ell>1$, the staggered spin-density correlations display a decreasing monotonic 
behavior, as shown in Fig.\ \ref{corr3}.
The ground state in this case is dominated by states such as
%\begin{equation}
$$|0\ \uparrow\ \, \downarrow\ \, \uparrow\ \, \downarrow\ \ \uparrow\downarrow
\  \ \uparrow\ \, \downarrow\ \, \uparrow\ \, \downarrow\ \, \uparrow\ \, \downarrow\rangle +
{\rm C.p.,}\nonumber\\
$$%\end{equation}
for $N_s=12$ and similarly for $N_s=18$. 
Note that the DO site is located in the ring exactly opposite to the empty site,
and that the alternating sequence of up and down spins is displaced by one 
site whenever one goes through either an empty or a DO site; this explains
the sign change in ${\cal F}_S(\ell)$ in Fig.\ \ref{corr3} near $\ell=N_s/4$. 
For comparison, we also show in Fig.\ \ref{corr3} the staggered 
spin-density correlation function 
in the absence of bond-charge interaction in the SDW region.
The ``kinked'' correlations obtained when $X=1$ are quite robust, and
their slower spatial decay is suggestive of some degree of ordering.
In the region where the ground state corresponds to 2 DO sites 
the behavior is qualitatively the same, and this may be a trademark
of the DO region.

The degeneracy of the state with $S^z=0$ with that having maximum $S^z$
is also manifested when comparing the spin-density correlation functions:
Apart from the slightly larger value at $\ell=0$ (i.e., essentially 
the local moment), the amplitude does not decay with the distance,
reflecting the strong pinning of the quasi-ferromagntic state. 

Similarly to $|X|<1$, superconducting correlations are not enhanced
by bond-charge interaction. 

\section{CONCLUSION}
\label{sec:concl}
The effects of bond-charge interaction on the extended
Hubbard model have been inferred from the analysis of a particular
gap function  to locate the transition points, and from
various correlation functions; 
in the absence of bond-charge interaction, i.e., $X=0$, this procedure
reproduces the known results quite accurately.  
For fixed $X\neq 1$, we find that this term causes 
the charge-density--wave region to grow at the expense of the 
spin-density--wave phase; 
this growth increases with $|X|$ and is more pronounced for $X>0$ 
than for $X<0$. 
As far as  correlations are concerned, bond-charge interaction 
smooths the CDW-SDW transition, possibly driving them to second order or,
at least, weakening their first order nature; this point is surely 
worth being pursued further.  

For $X=1$, we have presented evidence indicating that the intermediate
phase, between the charge-ordered and singly-occupied states, comprises 
a succession of ground states corresponding to a continuous variation
of the density of doubly-occupied sites. 
Further, these $S^z=0$ states are degenerate with those having maximum 
$S^z$ compatible with the density of doubly-occupied sites. 
Associated with these doubly-occupied states, the staggered spin-density
correlation functions corresponding to $S^z=0$ develop ``kinks'', 
while those corresponding to a maximum $S^z$ display pinning behavior.

As far as the possibility of bond-charge interaction favoring 
a superconducting state, we have found no enhancement of 
pairing correlations, for the values of $X$ considered here. 
This does not rule out the possibility of pairing for larger values 
of $|X|$, for higher dimensions
or for other band fillings; we are currently investigating the possibilities
of other fillings and higher dimensions.

Upon completion of this work, we received a preprint by Arrachea
et al.,\cite{Arrachea96} in which the model discussed in 
Refs.\ \onlinecite{Strack93,Ovchinnikov934,Aligia95,Arrachea94} is studied by several
methods; their model reduces to the one presented here in
some special cases, for which the overall results agree with ours.

\acknowledgments
We are grateful to F.~C.~Alcaraz, A.~Aligia, S.~S.~Makler, and T.~J.~P.~Penna
for very interesting suggestions.
We are also grateful to A. Aligia and L. Arrachea for sending
us several preprints prior to publication. 
The calculations were partly performed at the Centro Nacional de
Supercomputa\c c\~ao of the Federal University of Rio Grande do Sul.
Financial support from the Brazilian Agencies 
FINEP/MCT, 
%Minist\'erio de Ci\^encia e Tecnologia, 
CNPq, 
%Conselho Nacional de Desenvolvimento Cient\'\i fico e Tecnol\'ogico (CNPq) 
and CAPES 
%Coordena\c c\~ao de Aperfei\c coamento do Pessoal de Ensino Superior (CAPES), 
is also gratefully acknowledged.

\begin{figure}
%\epsfxsize=8truecm
%\centerline{\epsffile{vux.eps}}
\caption{Phase diagram $V\ vs.\ U$ for the usual (i.e., $X=0$) 
extended Hubbard model at half-filling. 
CDW and SDW denote charge- and spin-density--wave ground states.
The dotted line is the strong coupling prediction $V_c=U/2$, and 
the solid line guides the eye through our extrapolated results, shown
as solid circles. 
For comparison, some results from Ref.\ 11 are shown as empty circles.} 
\label{pd1} %fig1
\end{figure} 

\begin{figure}
%\epsfxsize=8truecm
%\centerline{\epsffile{cdw.eps}}
%\centerline{\epsffile{sdw.eps}}
\caption{Charge-density (circles) and spin-density (squares) 
correlation functions {\it vs.} intersite distance, for a
10-site chain with $V=2$, in (a) the CDW phase ($U=3.2$), and (b) the SDW
phase ($U=4.8$);
dotted lines and open symbols correspond to $X=0$ and 
solid lines and solid symbols to $X=-0.1$.}
\label{corr2} %fig2, formerly fig4
\end{figure}

\begin{figure}
%\epsfxsize=8truecm
%\centerline{\epsffile{vux.eps}}
\caption{Extrapolated phase diagram $V\ vs.\ U$ for the extended Hubbard 
model at half-filling. 
The solid line represents the system without bond-charge interaction, and 
the other lines correspond to $X=-0.1$ (dotted), $X=-0.5$ (dashed),
$X=0.5$ (long-dashed), and $X=-1$ (dot-dashed).}
\label{fphased} %fig3
\end{figure} 

\begin{figure}
%\epsfxsize=8truecm
%\centerline{\epsffile{fig2.eps}}
\caption{Extrapolated lowest energy levels per fermion for different 
densities of doubly occupied sites, $\rho_2\equiv N_2/N_s$, which label the
curves.}
\label{energies3} %fig4, formerly fig7
\end{figure}

\begin{figure}
%\epsfxsize=8truecm
%\centerline{\epsffile{vux1.eps}}
\caption{Phase diagram $V\ vs.\ U$ for the extended Hubbard model at
half-filling and $X=1$. CO, SO, and DO respectively stand for charge-ordered, 
singly-occupied and doubly-occupied states.} 
\label{PDX1} %fig5, formerly fig8
\end{figure}

\begin{figure}
%\epsfxsize=8truecm
%\centerline{\epsffile{cdw.eps}}
%\centerline{\epsffile{sdw.eps}}
\caption{Staggered spin-density correlation function vs. inter-site distance 
in the $S^z=0$ sector, for $U=2$ and $V=0.5$.
Short-dashed lines represent data for $X=0$ in the SDW region with $N_s=12$
sites.
Long-dashed (solid) lines represent data for $X=1$ with a ground state   
corresponding to one DO site on a chain with $N_s=12$ ($N_s=18$) sites.}
\label{corr3} %fig6, formerly fig9
\end{figure}
%\noindent

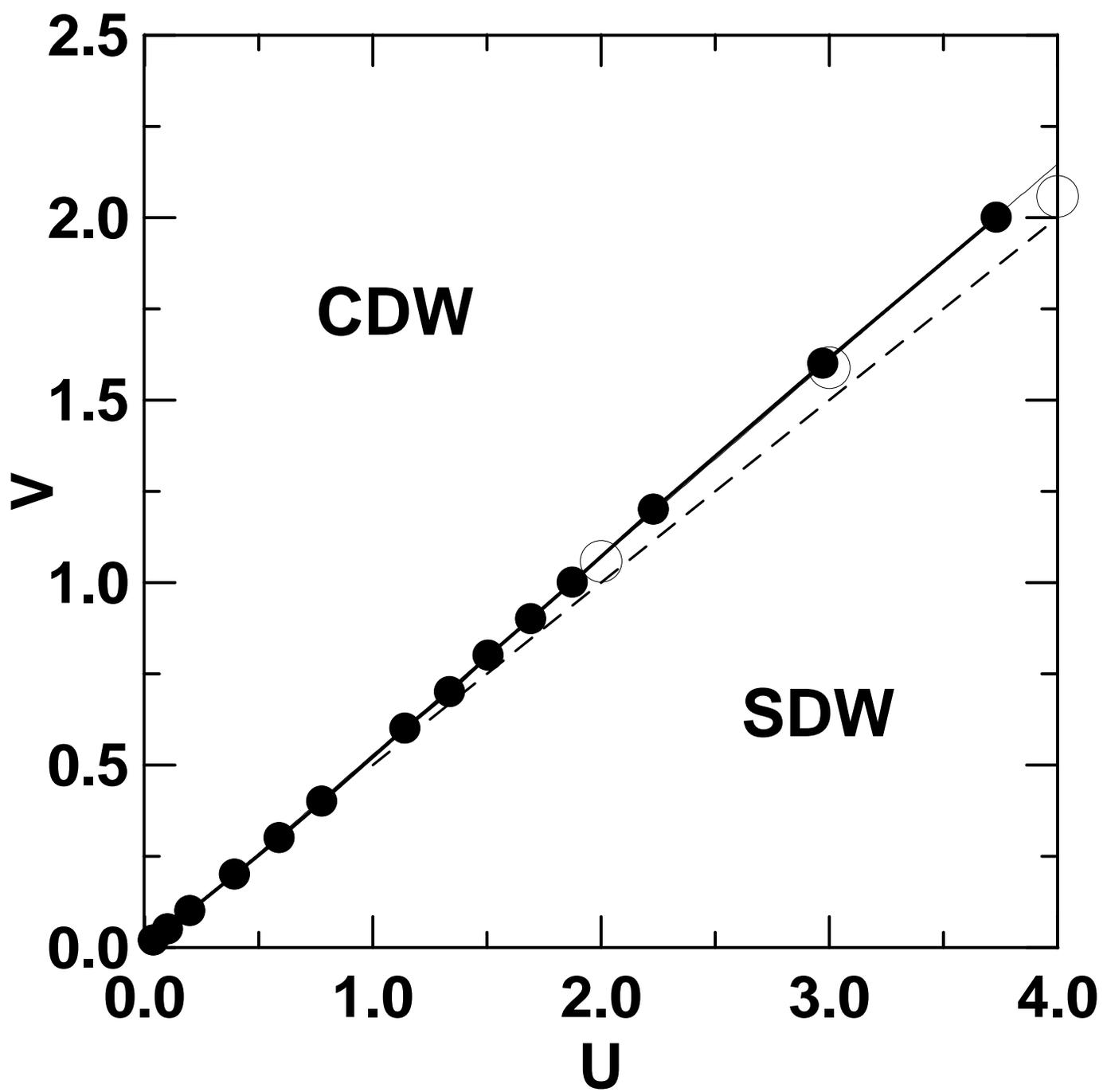
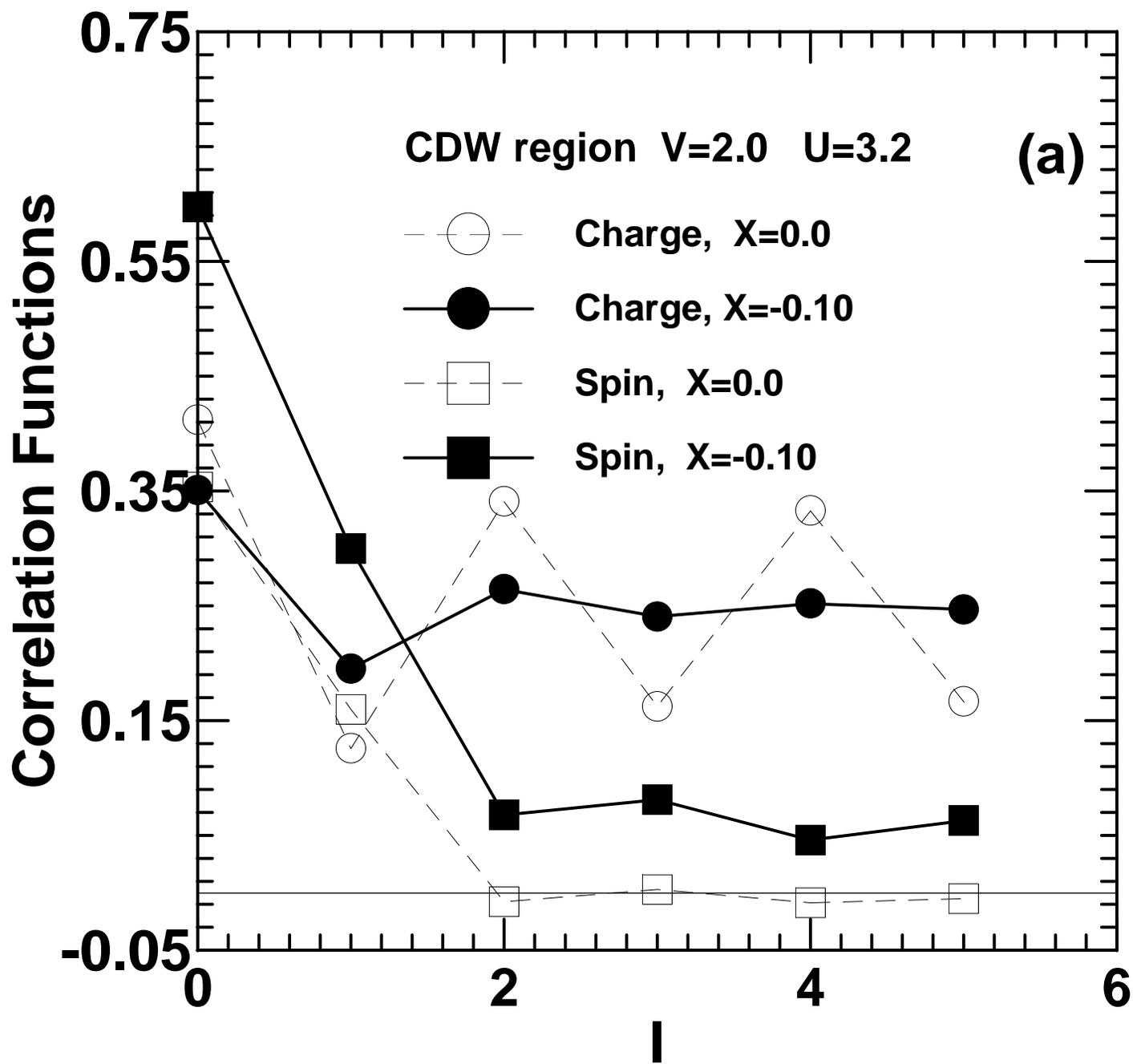
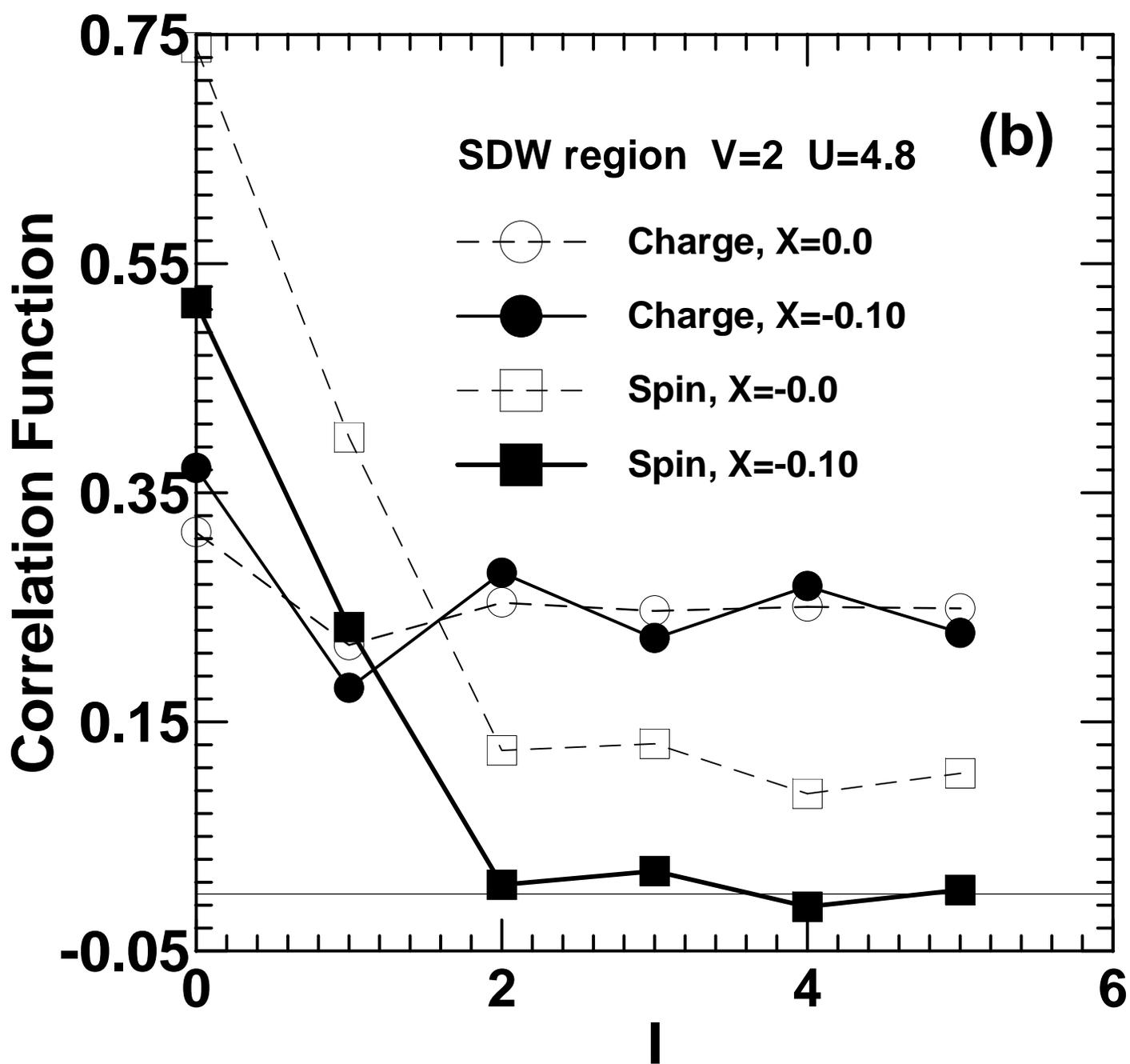
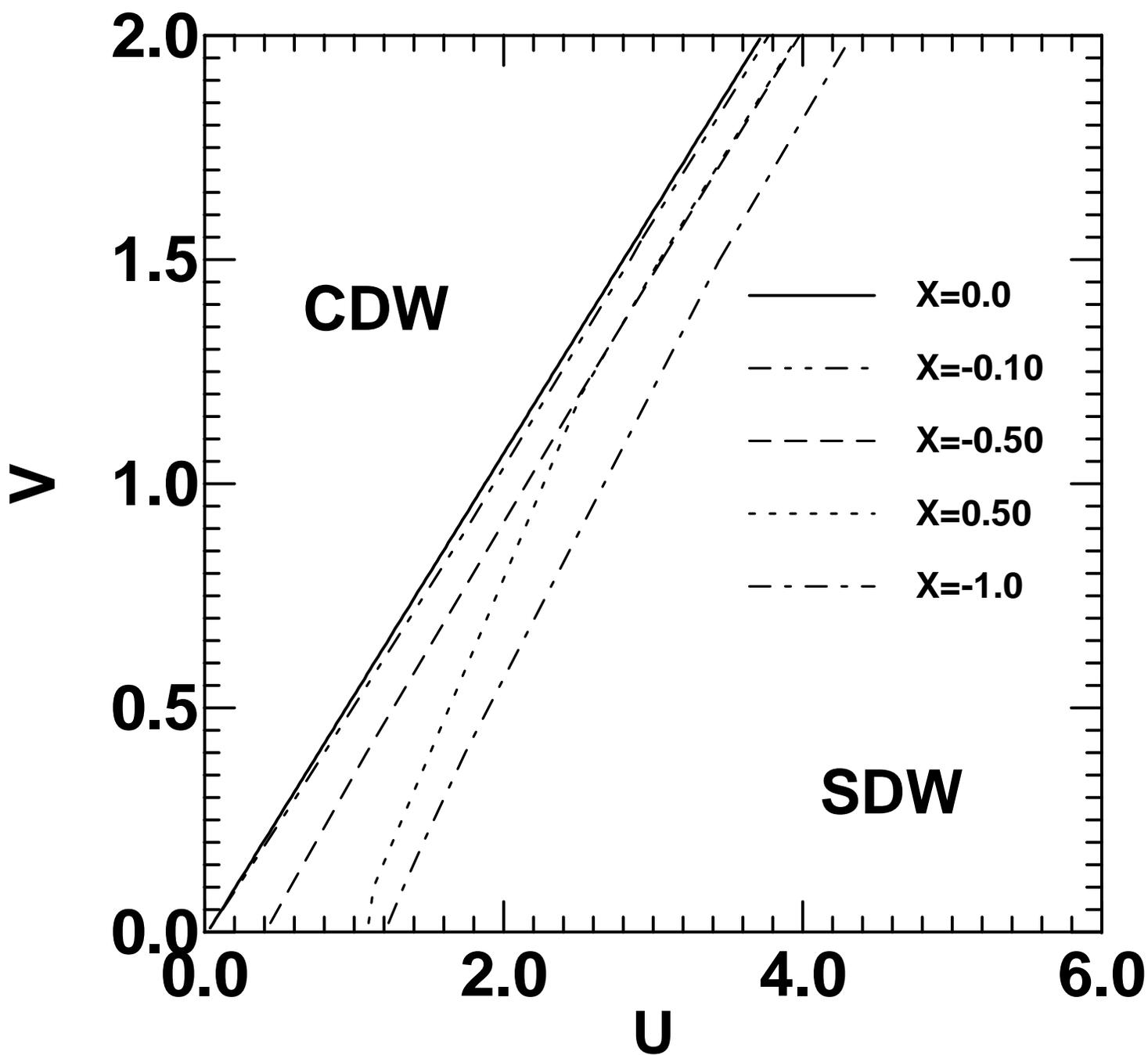
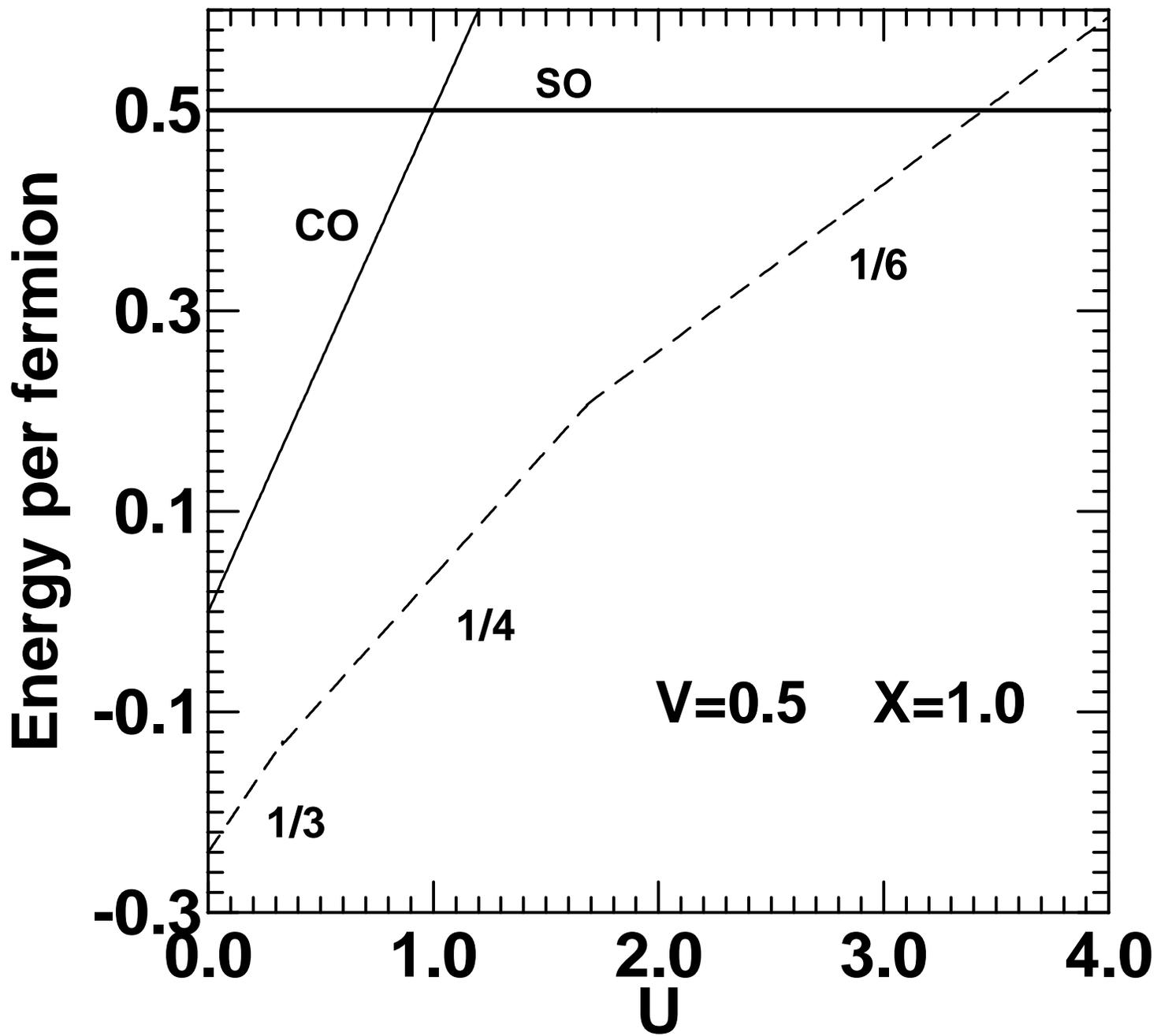
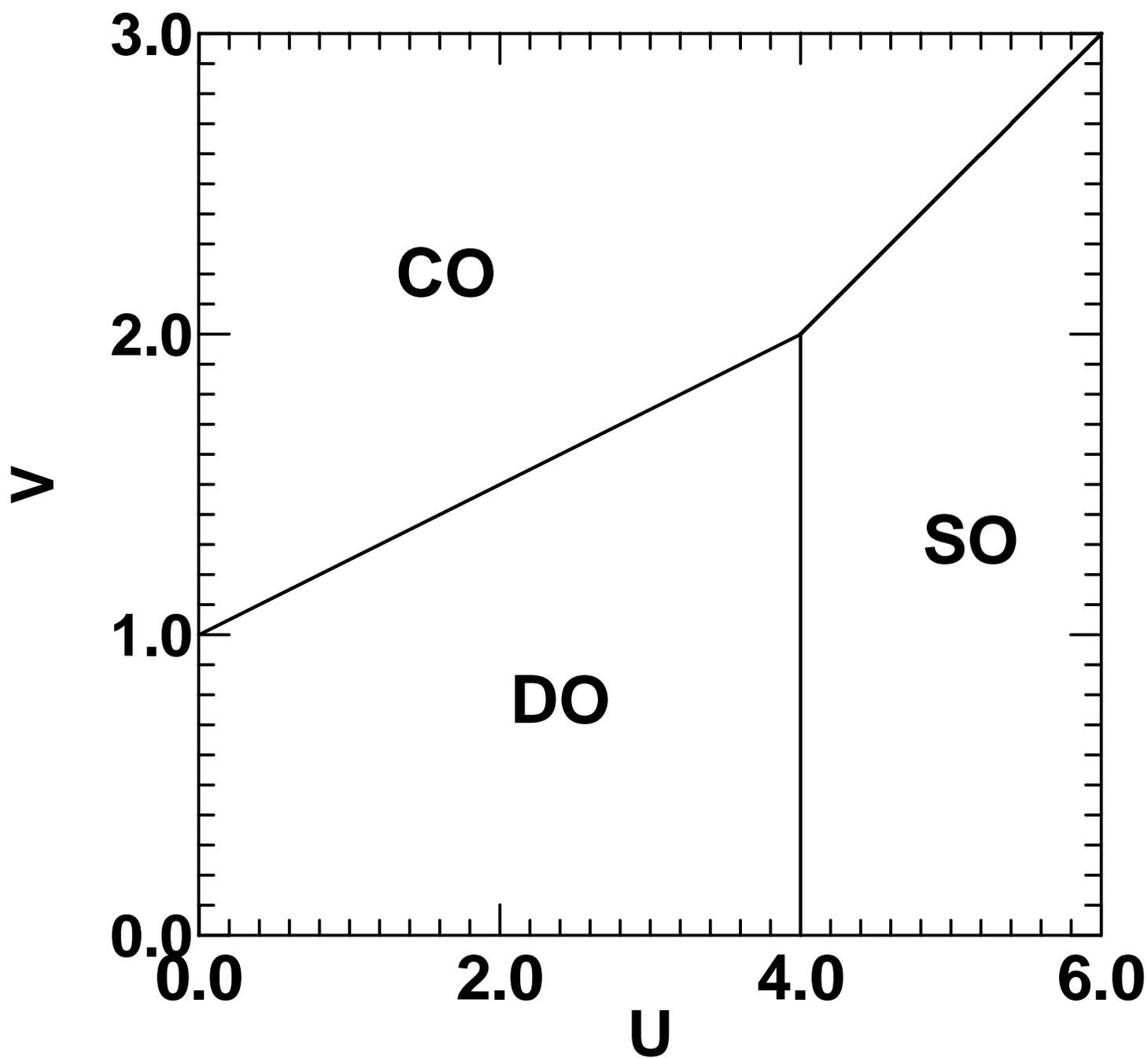
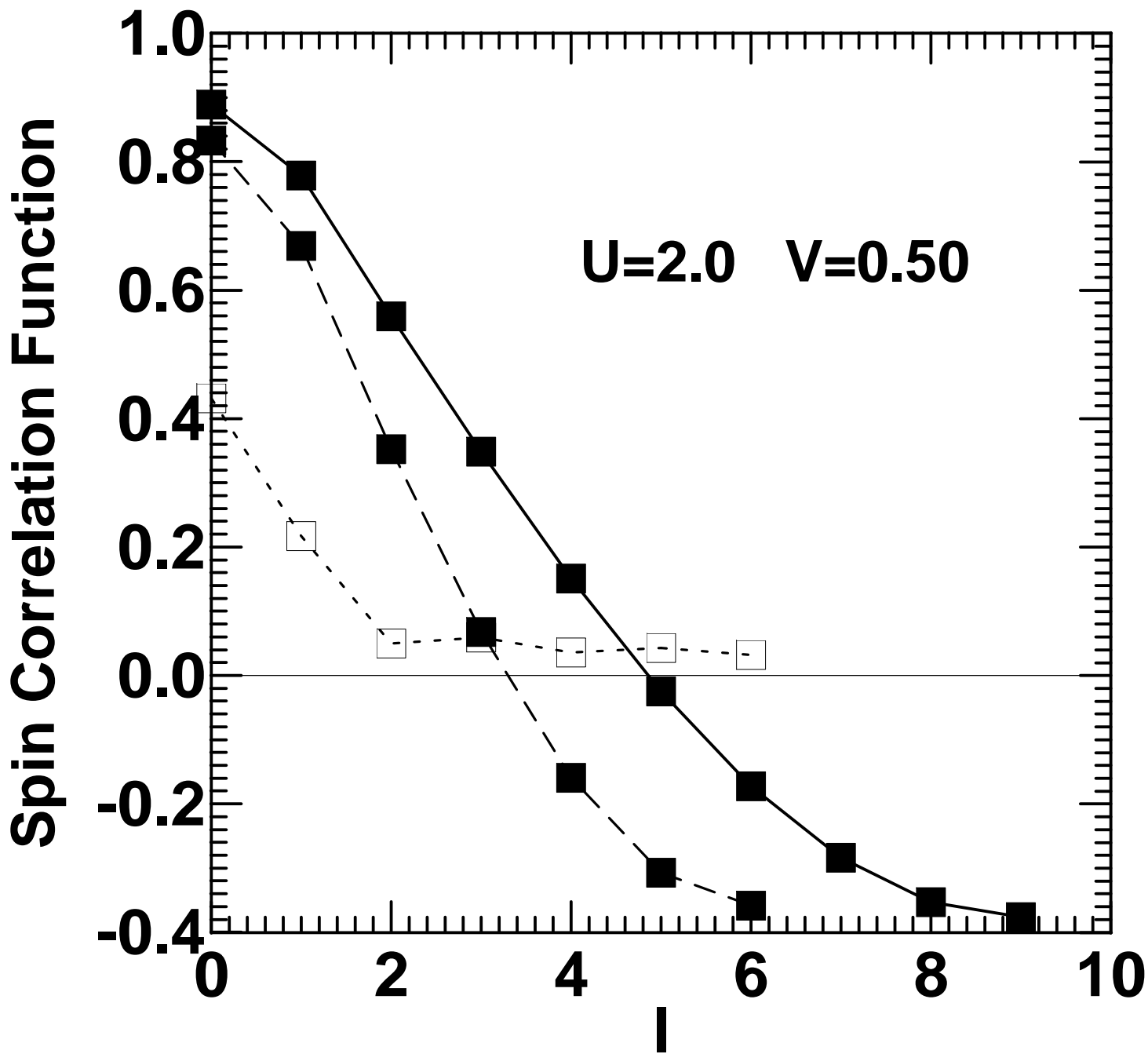
\end{document}